\documentclass[aps,prd,amsmath,amssymb,twocolumn,floatfix,superscriptaddress]{revtex4-1}
\usepackage{graphicx}
\pdfoutput=1
\usepackage{times}
\usepackage{multirow}
\usepackage{epstopdf}

\usepackage[margin=1in]{geometry}

\usepackage{amsmath}
\usepackage{graphicx}

\begin{document}
\null\hfill\begin{tabular}[t]{l@{}}
\small{FERMILAB-PUB-18-559-ND}
\end{tabular}

\author{R.~Acciarri}
\affiliation{Fermi National Accelerator Lab, Batavia, Illinois 60510, USA}

\author{C.~Adams}
\affiliation{Yale University, New Haven, Connecticut 06520, USA}

\author{J.~Asaadi}
\affiliation{University of Texas at Arlington, Arlington, Texas 76019, USA}

\author{B.~Baller}
\affiliation{Fermi National Accelerator Lab, Batavia, Illinois 60510, USA}

\author{T.~Bolton}
\affiliation{Kansas State University, Manhattan, Kansas 66506, USA}

\author{C.~Bromberg}
\affiliation{Michigan State University, East Lansing, Michigan 48824, USA}

\author{F.~Cavanna}
\affiliation{Fermi National Accelerator Lab, Batavia, Illinois 60510, USA}

\author{E.~Church}
\affiliation{Pacific Northwest National Lab, Richland, Washington 99354, USA}

\author{D.~Edmunds}
\affiliation{Michigan State University, East Lansing, Michigan 48824, USA}

\author{A.~Ereditato}
\affiliation{University of Bern, 3012 Bern, Switzerland}

\author{S.~Farooq}
\affiliation{Kansas State University, Manhattan, Kansas 66506, USA}

\author{A.~Ferrari}
\affiliation{CERN, CH-1211 Geneva 23, Switzerland}

\author{R.S.~Fitzpatrick}
\affiliation{University of Michigan, Ann Arbor, Michigan 48109, USA}

\author{B.~Fleming}
\affiliation{Yale University, New Haven, Connecticut 06520, USA}


\author{A.~Hackenburg}
\affiliation{Yale University, New Haven, Connecticut 06520, USA}


\author{G.~Horton-Smith}
\affiliation{Kansas State University, Manhattan, Kansas 66506, USA}

\author{C.~James}
\affiliation{Fermi National Accelerator Lab, Batavia, Illinois 60510, USA}

\author{K.~Lang}
\affiliation{University of Texas at Austin, Austin, Texas 78712, USA}

\author{M.~Lantz}
\affiliation{Uppsala University, 751 20 Uppsala, Sweden}

\author{I.~Lepetic}
\email{ilepetic@hawk.iit.edu}
\affiliation{Illinois Institute of Technology, Chicago, Illinois 60616, USA}

\author{B.R.~Littlejohn}
\email{blittlej@iit.edu}
\affiliation{Illinois Institute of Technology, Chicago, Illinois 60616, USA}

\author{X.~Luo}
\affiliation{Yale University, New Haven, Connecticut 06520, USA}

\author{R.~Mehdiyev}
\affiliation{University of Texas at Austin, Austin, Texas 78712, USA}

\author{B.~Page}
\affiliation{Michigan State University, East Lansing, Michigan 48824, USA}

\author{O.~Palamara}
\affiliation{Fermi National Accelerator Lab, Batavia, Illinois 60510, USA}


\author{B.~Rebel}
\affiliation{Fermi National Accelerator Lab, Batavia, Illinois 60510, USA}

\author{P.R.~Sala}
\affiliation{INFN Milano, INFN Sezione di Milano, I-20133 Milano, Italy}

\author{G.~Scanavini}
\affiliation{Yale University, New Haven, Connecticut 06520, USA}

\author{A.~Schukraft}
\affiliation{Fermi National Accelerator Lab, Batavia, Illinois 60510, USA}

\author{G.~Smirnov}
\affiliation{CERN, CH-1211 Geneva 23, Switzerland}

\author{M.~Soderberg}
\affiliation{Syracuse University, Syracuse, New York 13244, USA}

\author{J.~Spitz}
\affiliation{University of Michigan, Ann Arbor, Michigan 48109, USA}

\author{A.M.~Szelc}
\affiliation{University of Manchester, Manchester M13 9PL, United Kingdom}

\author{M.~Weber}
\affiliation{University of Bern, 3012 Bern, Switzerland}

\author{W.~Wu}
\affiliation{Fermi National Accelerator Lab, Batavia, Illinois 60510, USA}

\author{T.~Yang}
\affiliation{Fermi National Accelerator Lab, Batavia, Illinois 60510, USA}

\author{G.P.~Zeller}
\affiliation{Fermi National Accelerator Lab, Batavia, Illinois 60510, USA}

\collaboration{The ArgoNeuT Collaboration}
\noaffiliation

\title{Demonstration of MeV-Scale Physics in
\\Liquid Argon Time Projection Chambers Using ArgoNeuT}

\begin{abstract}
MeV-scale energy depositions 
by low-energy photons produced in neutrino-argon interactions have been identified and reconstructed in ArgoNeuT liquid argon time projection chamber (LArTPC) data. ArgoNeuT 
data collected on the NuMI beam at Fermilab
were analyzed to select 
isolated low-energy depositions in the TPC volume.
The total number, reconstructed energies and positions of these depositions have been compared to those from simulations of neutrino-argon interactions using the FLUKA Monte Carlo generator. Measured features are consistent with energy depositions from photons produced by de-excitation of the neutrino's target nucleus and by inelastic scattering of primary neutrons produced by neutrino-argon interactions. This study represents a successful reconstruction of physics at the MeV-scale in a LArTPC, a capability of crucial importance for detection and reconstruction of supernova and solar neutrino interactions in future large LArTPCs.  
\end{abstract}

\maketitle

\section{Introduction}
\label{sec:intro}
The Liquid Argon Time Projection Chamber (LArTPC) is a powerful detection technology for neutrino experiments, as it allows for millimeter spatial resolution, provides excellent calorimetric information for particle identification, and can be scaled to large, fully active, detector volumes. LArTPCs have been used to measure neutrino-argon interaction cross sections and final-state particle production rates in the case of ArgoNeuT~\cite{CCinclNeutrino,flux, hammer,coherent,eventselectionpaper,pi0,CC1pion} and MicroBooNE~\cite{cpm}, neutrino oscillations in the case of ICARUS~\cite{icarusosc}, and charged particle interaction mechanisms on argon in the case of LArIAT~\cite{lariat}. 

LArTPCs are being employed to make important measurements, e.g. understanding the neutrino-induced low-energy excess of electromagnetic events with MicroBooNE~\cite{microboonedetector} and will be used to search for sterile neutrinos in the Fermilab SBN program~\cite{sbnd} and for CP-violation in the leptonic sector with DUNE~\cite{dunephysics}. Precise measurements of neutrino-argon cross sections will be performed with SBN~\cite{sbnd} and of charged hadron interactions with ProtoDUNE~\cite{protodunetdr}.
In most of the existing measurements, LArTPCs were placed in high energy neutrino beams to study GeV-scale muon and electron neutrinos as well as final-state products, generally with energies greater than $100$ MeV. A smaller number of measurements have investigated particles or energy depositions in the $<100$ MeV range~\cite{microboonemichel, icarusdecay, pi0}, some using scintillation light~\cite{lariatscint}.  

Few existing measurements have demonstrated LArTPC capabilities at the MeV scale for neutrino experiments, despite the wealth of physics studies that have been proposed for future large LArTPCs in this energy range. A number of studies have investigated expected supernova and solar neutrino interaction rates in the DUNE experiment: see Refs. ~\cite{dunephysics} and ~\cite{scholberg2012supernova} for reviews and relevant citations. Other studies have proposed using decay-at-rest neutrino interactions for short-baseline oscillation tests, coherent neutrino scattering measurements and supernova-related studies~\cite{dar,captain,spitz,Akimov:2013yow,Brice:2013fwa}. LArTPC experiments utilizing GeV-scale neutrino beamlines would also benefit from the ability to perform a reconstruction of MeV-scale features. This ability would allow for a fuller reconstruction of beam neutrino events by enabling reconstruction of photons released during de-excitation of the 
nucleus and of part of the energy transferred to final-state neutrons. 
Furthermore, MicroBooNE has shown that identifying and including full reconstructed energies at ends of showers is challenging and would benefit from the ability to reconstruct Compton scatters of photons exiting the shower core~\cite{microboonemichel}. 

Performing identification and reconstruction of particles at MeV energies in a LArTPC is a challenging task. At higher energies ($>100$ MeV), charged particles travel several centimeters to meters in distance, leaving detectable signals on dozens to hundreds of TPC wires, producing an ionization track that can be utilized for reconstructing the identity and kinematics of detected particles. On the other hand, charged particles with kinetic energies near the MeV scale travel a distance of the order of or less than the distance between adjacent wires in many LArTPCs ($3$-$5$ mm), leaving just one hit or a short cluster of a few consecutive hits. Thus, current analysis methods used to reconstruct physics quantities from tracks made of large numbers of wire signals are ineffective in this energy regime, and there is a need for new, low-energy-specific methods.

We have used data acquired by the ArgoNeuT LArTPC detector at Fermilab to search for small energy depositions associated with neutrino events and compared them to predictions from the FLUKA neutrino interaction generator~\cite{fluka,fluka2,NUNDIS}. Using new topological reconstruction tools, we find clear evidence of activity due to de-excitation of the final-state nucleus and inelastic scattering of neutrons in the detector. 


We begin with a  description of the ArgoNeuT detector in Section \ref{sec:detector}. We then overview nuclear de-excitation photon production, photon emission from inelastic scattering of neutrons, and photon propagation in argon in Section \ref{sec:theory}. We then describe utilized datasets and reconstruction in Sections \ref{sec:datasets} and \ref{sec:signalselection}. Final reconstructed signal distributions are presented and compared to a Monte Carlo (MC) simulation in Section \ref{sec:data}.

\section{The ArgoNeuT Detector}
\label{sec:detector}
ArgoNeuT was a LArTPC experiment which was placed in the Neutrinos at the Main Injector (NuMI) beamline at Fermilab for five months in 2009-2010. ArgoNeuT was located $100$ m underground, in front of the MINOS near detector (MINOS ND). The TPC was $47 (\textit{w}) \times 40 (\textit{h})  \times 90 (\textit{l})$ cm$^3$ with a volume of $169$ L. Ionized charge drifted in the x-direction by means of an electric field produced by a cathode biased at a negative high voltage of magnitude $23.5$ kV. A field shaping cage caused the electric field along the drift length to be uniform at $481$ V/cm. The resulting drift velocity was $1.57$ mm/$\mu$s, with a maximum drift time of $300.5 \, \mu$s. At the anode end of the TPC there were three wire planes, of which two were instrumented (the innermost plane was a shield plane). The middle wire plane was the induction plane; the outer one was the collection plane. Each of the instrumented planes was comprised of $240$ wires, with a wire spacing of $4$ mm and oriented at $\pm 60^{\circ}$ to the beam direction. In each detector readout, each wire channel was sampled every $198$ ns, for a total readout window of $405 \, \mu$s. The waveform for each wire was recorded with hits identified from peaks above baseline. Triggering for a readout was determined by the NuMI beam spill, at a rate of $0.5$ Hz. A more detailed description and operational parameters of the ArgoNeuT detector are given in ~\cite{main}. 

ArgoNeuT benefited from the presence of the MINOS ND located immediately downstream of it. The MINOS ND is a segmented magnetized steel and scintillator detector~\cite{minos}. As a result, the momenta and signs of muons produced by neutrino interactions in ArgoNeuT and entering the MINOS ND could be determined by using reconstruction information from the MINOS ND. ArgoNeuT also benefited from its placement $100$ m underground; at this depth, cosmic rays are expected to be seen in fewer than $1$ in $7000$ triggers. 

During the majority of ArgoNeuT's run, the NuMI beam was operated in the low energy antineutrino mode; neutrino fluxes produced during this operation mode are described in~\cite{flux}. The composition of the beam was $58$\% muon neutrino, $40$\% muon antineutrino, and $2$\% electron neutrino and antineutrino. The average energy for muon neutrinos was $9.6$ GeV, and the average energy of muon antineutrinos was $3.6$ GeV. The antineutrino mode run lasted $4.5$ months with $1.25 \times 10^{20}$ protons on target (POT) acquired.

\section{Production and interaction of low-energy photons in neutrino-argon interactions}
\label{sec:theory}
MeV-energy photons can be produced in neutrino-argon interactions by two possible mechanisms, de-excitation of the target nucleus and inelastic scattering of final-state particles. When a neutrino interacts with an $^{40}$Ar nucleus, the target nucleon and the neutrino interaction products initiate a nuclear reaction during which nucleons and nuclear fragments may be emitted. The remaining residual nucleus is  often left in an excited state. The nucleus de-excites by means of the emission of a photon or cascade of photons with energies ranging from $\sim 0.1$ MeV -- $10$ MeV. Reaction products heavier than deuterons and the recoiling residual nucleus are generally not observable in a LArTPC.
Final-state neutrons which inelastically scatter off an $^{40}$Ar nucleus or are captured by it will also produce photons in the energy range of interest as the $^{40}$Ar nucleus de-excites~\cite{nndc}.  

As photons are neutral particles, they cannot be detected directly. Instead we detect electrons resulting from a photon interaction. The scale of the distance between subsequent energy depositions for one photon is given by the radiation length ($X_0$), which in liquid argon is $14$ cm. Over the $\sim 0.1$ -- $10$ MeV range of interest in this study, the most probable interaction process for photons in LAr is Compton scattering. In Compton scattering at this energy, each photon has a high probability of creating multiple topologically isolated energy depositions within a LArTPC. Higher energy photons can also interact via pair-production, however this is still subdominant in the energy range considered here.  

\subsection {Neutrino interactions and neutron scattering in FLUKA}
The only neutrino MC interaction generator that includes the simulation of both mechanisms of low-energy photon production in GeV-scale neutrino interactions in argon is FLUKA ~\cite{fluka,fluka2,NUNDIS}. FLUKA is a multi-particle transport and interaction code. Its neutrino interaction generator, called NUNDIS~\cite{NUNDIS}, is embedded in the same nuclear reaction module of FLUKA used for all hadron-induced reactions.  Quasi elastic, resonant ($\Delta$ production only), and deep inelastic scattering  interactions are modeled on single nucleons according to standard formalisms. Initial state effects are accounted for by considering bound nucleons distributed according to a Fermi momentum distribution. Final-state effects include a generalized intranuclear cascade (G-INC), followed by a pre-equilibrium stage and an evaporation stage. As mentioned above, nucleons, mesons and nuclear fragments can be emitted during these stages. Residual excitation is dissipated through photon emission. Experimental data on nuclear levels and photon transitions are taken into account whenever available.

Neutron-induced reactions are treated as standard hadronic interactions for neutron energies above 20~MeV, while for energies below 20~MeV a data-driven treatment is used, as in most low-energy neutron transport codes. Reaction cross sections, branching ratios and emitted particle spectra are imported from publicly available databases. Transport is based on a multi-group approach (neutron energies grouped in intervals, cross sections averaged within groups), except for selected reactions \cite{fluka}. In the FLUKA version used for this work (FLUKA2017, not yet released), a special treatment has been implemented for reactions on $^{40}$Ar. Cross sections are evaluated point-wise (for the exact neutron energy), correlations among reaction products are included, and gamma de-excitation is simulated as a photon cascade following experimental energies and branching ratios. 

\begin{figure}[t]
\includegraphics[scale=0.44, trim = 0.2cm 0.1cm 0.1cm 0.8cm, clip]{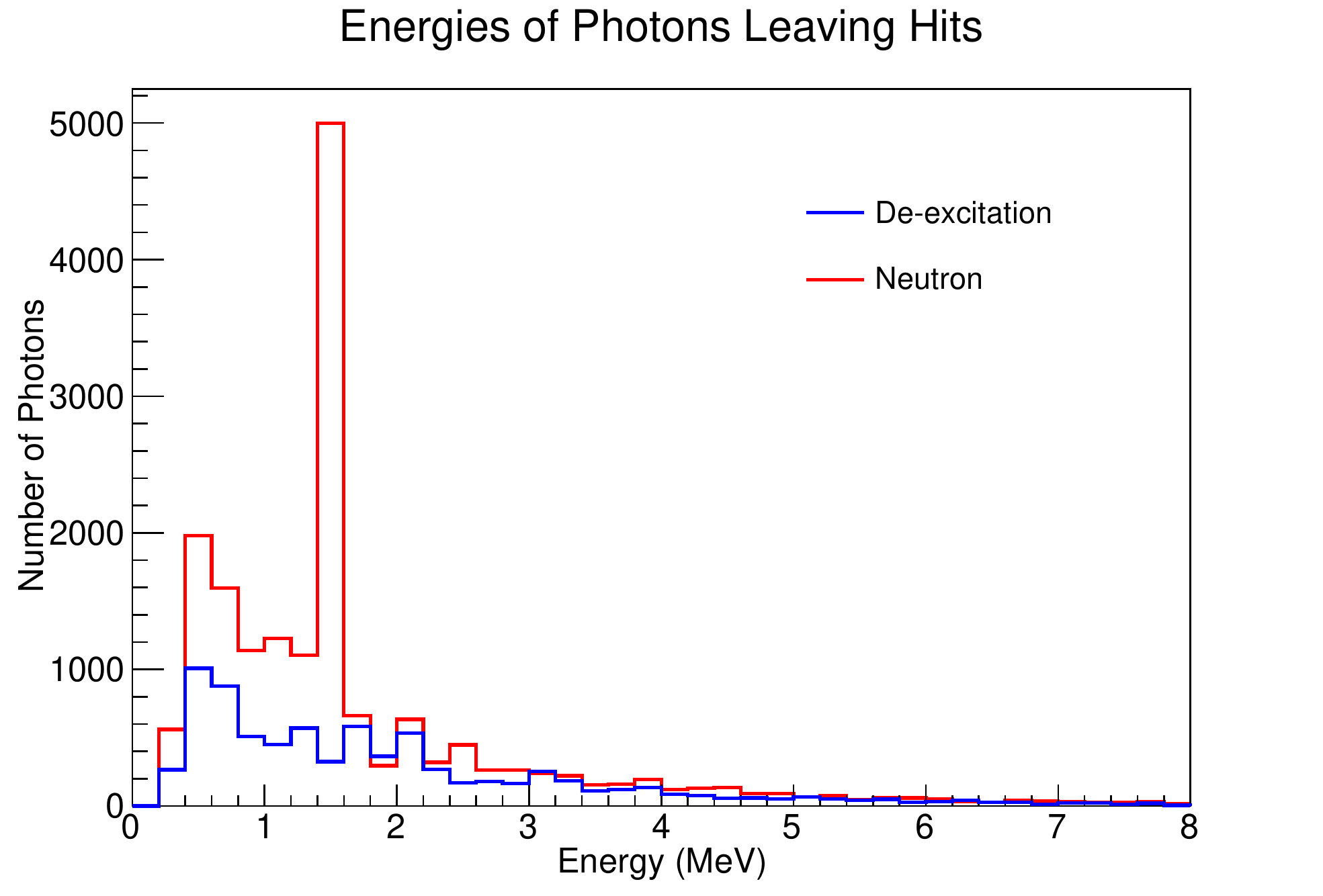}
\includegraphics[scale=0.44, trim = 0.2cm 0.1cm 0.1cm 0.8cm, clip]{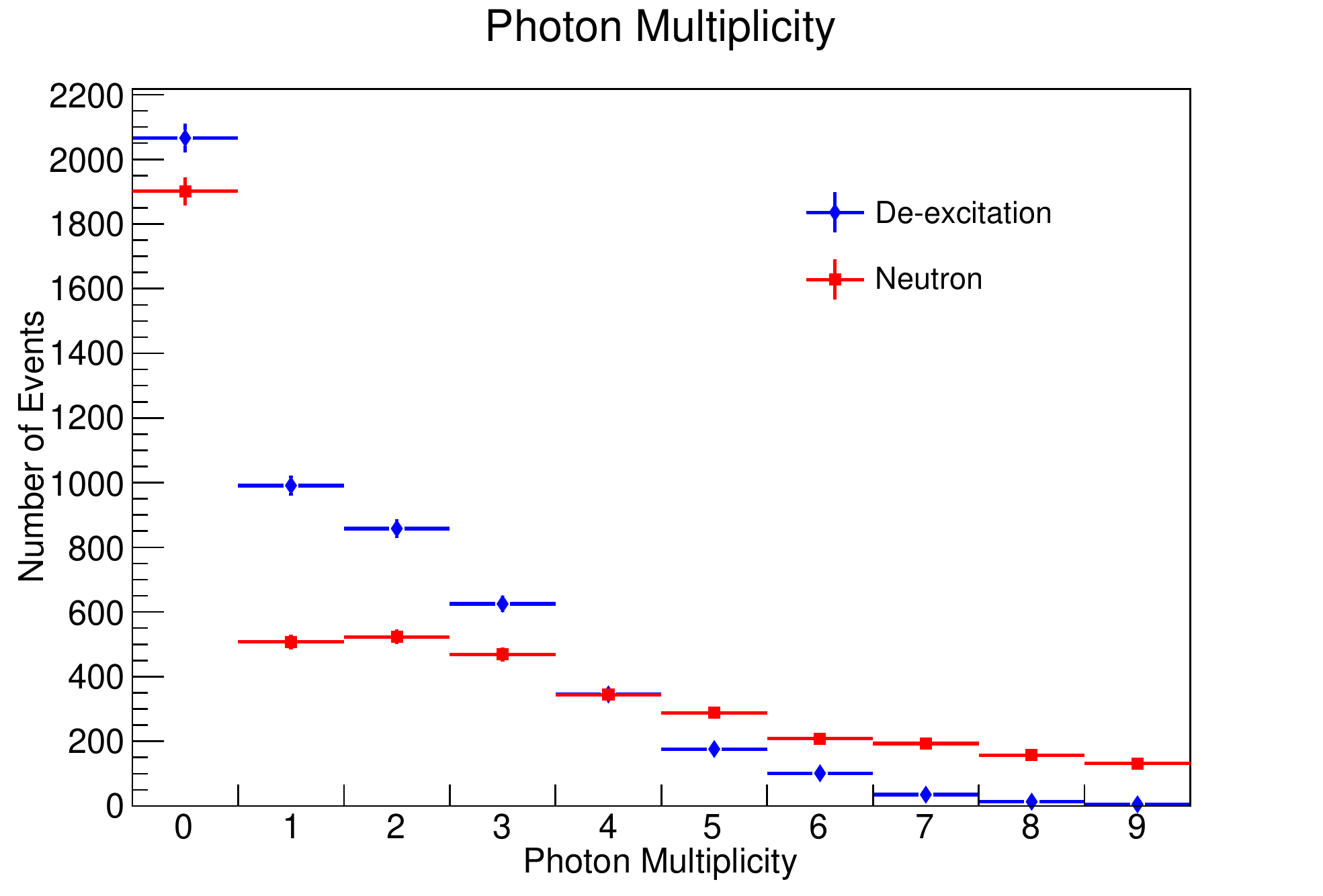}
\caption{Energy (top) and multiplicity (bottom) of low-energy photons from charged current interactions of muon neutrinos from the NuMI beam interacting and depositing energy in a volume of liquid argon with the dimensions of ArgoNeuT. Color indicates source of photon (blue are de-excitation photons, red are photons produced by neutrons). For a photon to be tracked in the simulation, it must have an energy $\geq 0.2$ MeV. The peak at $1.46$ MeV corresponds to the first excited state of $^{40}$Ar.}
\label{fig:neutronvsvertexN}
\end{figure}

Figure \ref{fig:neutronvsvertexN} shows the energies and numbers of photons from charged current interactions of muon neutrinos from the NuMI beam interacting and depositing energy in a volume of liquid argon with the dimensions of ArgoNeuT, according to FLUKA simulation (see Section \ref{sec:datasets} for details). A significant overlap in both the energies and numbers of photons from the two processes (de-excitation of the target nucleus and inelastic neutron scattering) is visible, making separation of the source of energy depositions difficult based on these metrics alone. Considering ArgoNeuT's size, a photon could leave the TPC with a significant amount of its energy undetected. It is also notable that $24$\% of product nuclei in this simulation are found in the ground state and produce no photons.

\begin{figure}[h]
\includegraphics[scale=0.55, trim = 0.2cm 0.1cm 0.1cm 0.1cm, clip]{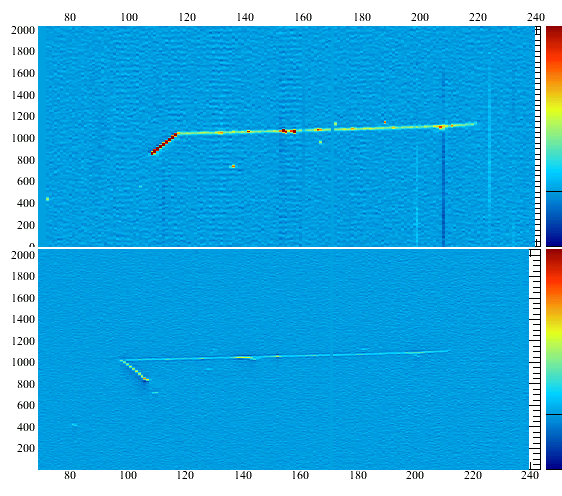}
\caption{A neutrino event (raw data) with one (longer) track reconstructed as a muon exiting the detector and one (shorter) track reconstructed as a proton. Possible photon activity (isolated blips) is visible in the event (e.g. collection plane wire 135, sample 700). The top image is the collection plane, and the bottom image is the induction plane. Wire number is indicated on the horizontal axis. The vertical axis indicates time sample number. Color indicates amount of charge collected.}
\label{fig:raw2}
\end{figure}

Typically, low energy photon-produced electrons are expected to appear in a LArTPC event display as blips from isolated energy depositions around the neutrino interaction vertex. An example can be seen in Fig. \ref{fig:raw2}, where a typical ArgoNeuT neutrino event is shown.


\section{Datasets}
\label{sec:datasets}
This analysis uses two primary real datasets from the antineutrino mode run. Events with simple, low track multiplicity final-state topology have been selected for the present analysis, as complex events make the selection of isolated low-energy signatures more difficult. The first dataset, termed the neutrino dataset, is a subsample of
muon neutrino and antineutrino 
events from the ArgoNeuT charged current pion-less (CC 0$\pi$) events sample, i.e. muon (anti)neutrino charged current events that do not produce pions in the final state. The selection and analysis of these events~\cite{eventselectionpaper}, requires that a three dimensional (3D) track reconstructed in the LArTPC is matched to a MINOS ND muon track, and that any number of tracks at the vertex, identified as protons using the algorithm defined in ~\cite{main}, are present in the final state ($\mu+Np$ events). In addition, we require that none of the events contains a reconstructed 3D track identified as a charged pion or a reconstructed shower corresponding to a high-energy electron or photon. The threshold for proton (pion) identification is $21$ ($10$) MeV~\cite{hammer}. From the CC 0-pion sample we have selected a subsample of events with one muon and up to one proton in the final state (CC 0$\pi$, 0 or 1 proton events) for the present analysis. 
The second dataset, termed the background dataset, was obtained by examining 
``empty event'' triggers which do not appear to contain a neutrino interaction. These readouts do contain ambient gamma ray activity, intrinsic $^{39}$Ar activity, photons produced by entering neutrons from neutrino interactions occurring upstream of the detector, and electronics noise. The beta emitter $^{39}$Ar is a radioactive isotope found in natural argon; at a rate of $1.38$ Bq/L, it is not expected to be a large background in ArgoNeuT events. Electronics noise can be identified as a hit if the deviation from the baseline is above a threshold. These features are also present in the neutrino events previously described, so the background dataset is used for a data-driven modeling of the background in the selected neutrino events.

ArgoNeuT data are compared with a MC dataset. We produced simulated neutrino interactions in ArgoNeuT using FLUKA and the energy spectrum of the NuMI beamline. A simplified ArgoNeuT detector geometry was inserted into FLUKA. 
In addition to producing all the final-state particles emerging from the neutrino interaction, including hadron re-interaction inside the nucleus (nuclear effects), FLUKA also simulates the physics of the final-state nucleus, resulting in the production of final-state de-excitation photons. 
FLUKA was also used to propagate final-state neutrons inside the LAr volume, resulting in the simulation of energies and locations of secondary neutron-produced photons. The FLUKA-determined properties of non-neutron final-state particles and secondary neutron-produced photons were then used as input to a LArSoft~\cite{larsoft} MC simulation of ArgoNeuT and propagated through the detector simulation, signal processing, and 
reconstruction stages as for real data. 
CC $0\pi$ $0,1$ proton events, i.e. events with one muon track entering the MINOS ND and up to one additional proton with kinetic energy $> 21$ MeV and no pions with kinetic energy $> 10$ MeV in the final state, compose the selected MC samples for the present analysis. Electronics noise, ambient and internal radioactivity, and photons from entering neutrons were not simulated; the background dataset described above was instead used to directly include these contributions to the MC dataset.

\section{Event Reconstruction}
\label{sec:signalselection}
As discussed in Section \ref{sec:theory}, 
the radiation length in liquid argon is $14$ cm, and MeV photon-produced electrons have ranges of a millimeter to a centimeter, as shown in Fig. \ref{fig:energyrange}. Consequently, for the present analysis a signal on the wire planes consists of a single hit or a very short cluster of hits on consecutive wires on both active planes of the TPC, topologically isolated from the rest of the event's features, possibly concentrated around the interaction vertex, as shown in Fig. \ref{fig:raw2}. 

The same reconstruction procedure has been applied to all the selected data and MC samples described in the previous Section. The reconstruction proceeded through two steps, one ``standard'' reconstruction step, followed by a low-energy specific second step, described in Section \ref{sec:photonreco}.

First, the ``standard'' ArgoNeuT automated reconstruction procedure, including hit finding, hit reconstruction and track reconstruction, as described in detail in ~\cite{CC1pion}, was applied. Events were required to have a reconstructed neutrino interaction vertex contained in the fiducial detector volume, defined as $[3,44]$ cm along the drift direction, $[-16,16]$ cm vertically from the center of the detector, and $[6,86]$ cm along the beam. The neutrino and background datasets contain $552$ and $1970$ events, respectively.

\subsection{Signal Selection}
\label{sec:photonreco}
In the second step, a low-energy specific procedure to identify and reconstruct isolated hits and clusters 
was applied. Since low-energy electrons will leave short isolated features in the TPC, hits that are identified as belonging to a reconstructed track longer than $1.5$ cm and beginning at the neutrino interaction vertex
were removed. To also remove nearby wire activity associated with a track (such as delta rays), all hits inside a $120^\circ$ cone around the first $2.4$ cm of each reconstructed track and a $5$ cm cylinder along the remaining track length were rejected. For tracks reconstructed as being longer than $4$ cm, the cylindrical rejection region was extended past the end of the track, in case the automated reconstruction cuts the track short. 

\begin{figure}[t]
\includegraphics[scale=0.45, trim = 0.5cm 0.2cm 0.3cm 1.1cm, clip]{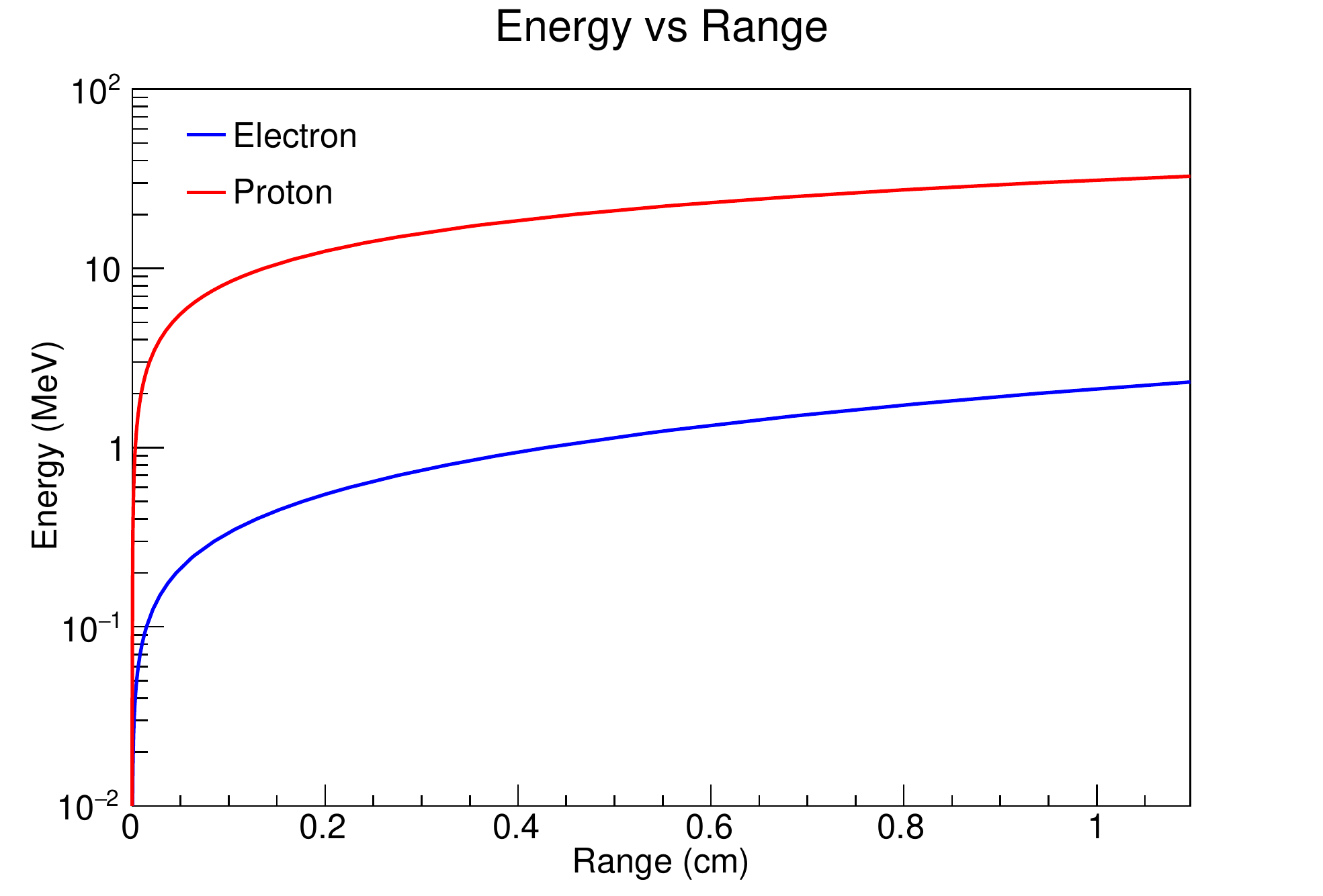}
\caption{Energy vs range for electrons and protons for the ranges of interest for this study. Red denotes protons, blue denotes electrons. The clear separation between electron and proton means it is unlikely a proton hit will be mistakenly identified as an electron hit. Data from ~\cite{nist}.}
\label{fig:energyrange}
\end{figure}

\begin{figure*}[!t]
\includegraphics[scale=0.40,width=0.45\textwidth, trim = 0.1cm 0.1cm 0.1cm 0.1cm, clip]{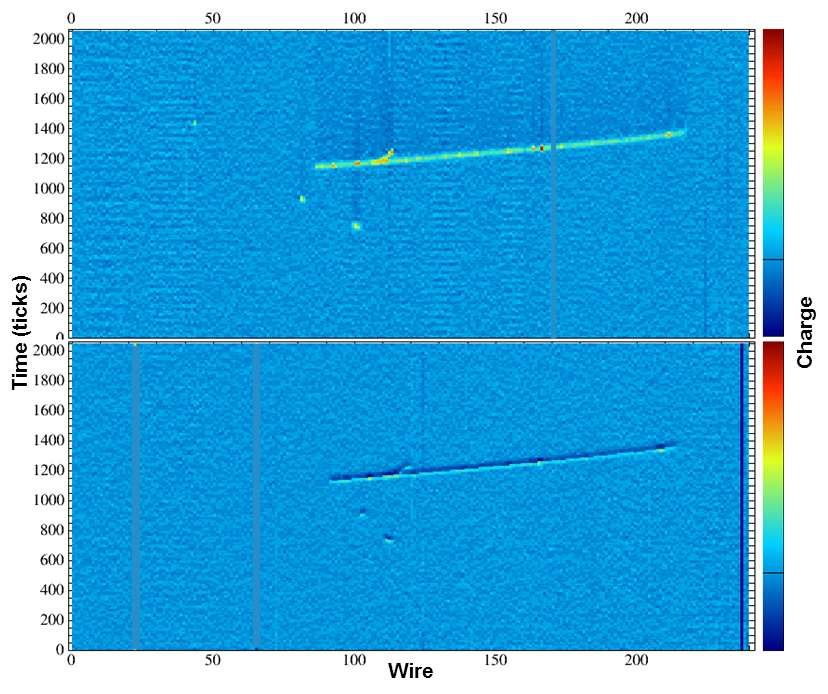}
\qquad \qquad
\includegraphics[scale=0.4,width=0.45\textwidth,trim = 0.2cm 0.1cm 1.0cm 0.1cm, clip]{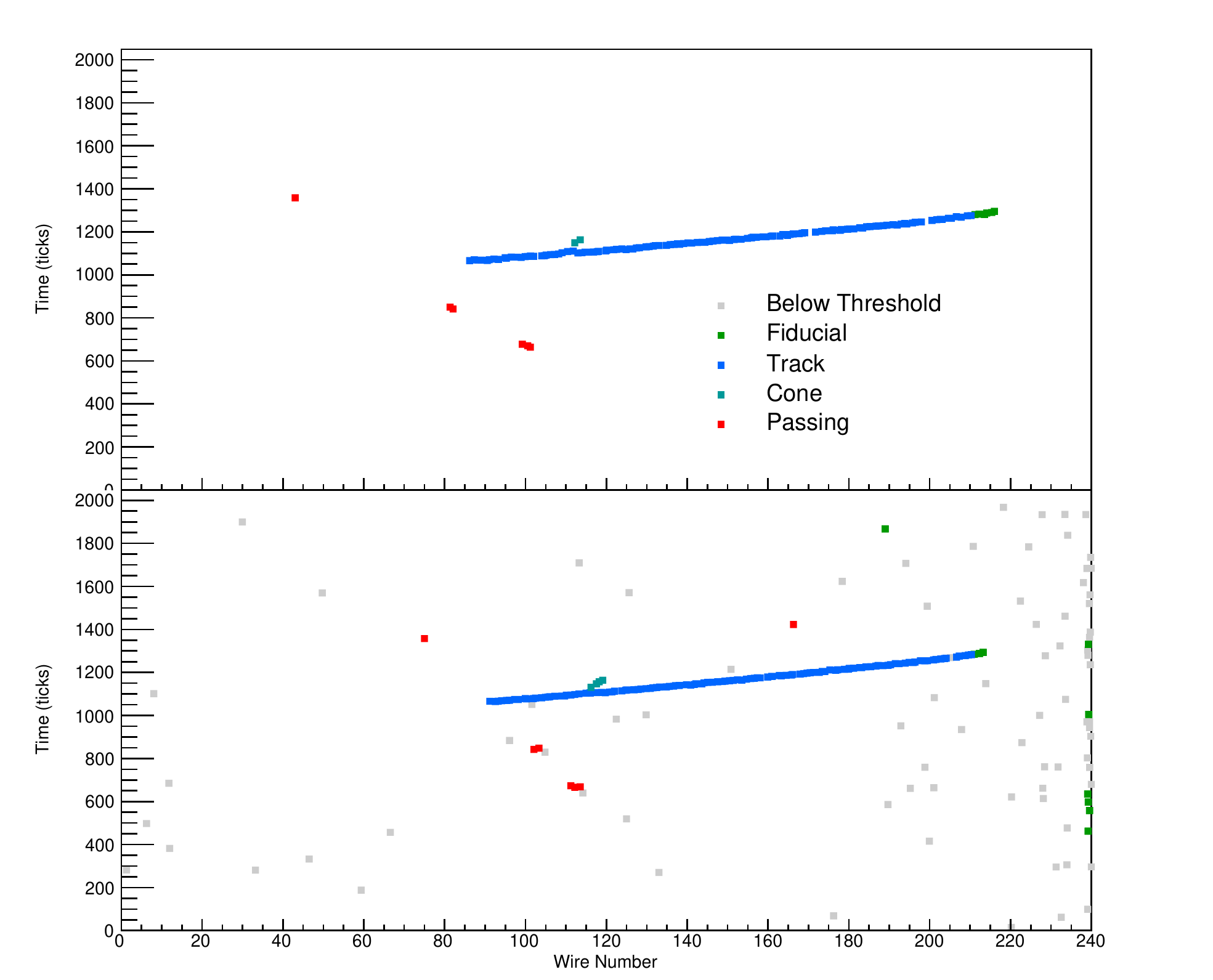}
\caption{Left: A raw data neutrino event display with one track reconstructed as a muon and with photon activity (isolated blips). The top image is the collection plane, and the bottom image is the induction plane. Wire number is indicated on the horizontal axis. The vertical axis indicates time sample number. Color indicates amount of charge collected. Right: The same event after hit finding and reconstruction. Each square denotes a reconstructed hit. Color indicates whether or not a hit was removed and by which cut (see text). Hits that pass all cuts are in red.}
\label{fig:hitsevd}
\end{figure*}

Then, several cuts were made on the remaining hits found in each event. A threshold cut removed hits whose fitted peak height is below 
a certain ADC count threshold on the induction and collection planes ($6$ and $10$ ADC, respectively), corresponding to roughly $0.2$ MeV of energy deposited. Hits whose fitted peak height is above a maximum ADC count ($60$ ADC, corresponding to $\sim 1.2$ MeV) were also removed, as they were unlikely to be produced by photon energy depositions. As shown in Fig. \ref{fig:energyrange}, such hits are more likely due to protons. For example, for a proton to travel a distance of $0.4$ cm, the wire spacing, it must have a kinetic energy of at least $21$ MeV, well above the maximum ADC cut. On the other hand, an electron must have a kinetic energy of $1$ MeV to travel the same distance. Low energy protons with very short range can result from a neutron-proton reaction on argon, however the FLUKA simulation indicates fewer than $1\%$ of hits passing cuts are due to protons. A fiducial cut was then applied to remove all hits within $6$ cm of the cathode and anode and hits near corners of the TPC. Real and MC events were individually visually scanned to remove noisy wires and reconstruction failures. Individual wires were removed on an event by event basis if it was clear they had several hits due to electronics noise, with equivalent cuts applied to background events. Some hits were also manually removed if it was clear they belonged to a track that was not reconstructed properly. To suppress hits originating from above-threshold electronics noise, matching of hit times between induction and collection planes was required. This plane matching also allowed for reconstruction of the 3D space position for all hits in the final sample passing the above selection criteria. Applied cuts are visually demonstrated in Fig. \ref{fig:hitsevd}.

A summary of the level of hit removal achieved in each cut for neutrino, background and MC datasets is found in Table \ref{table:cuts}. Once all cuts were applied and visual scanning was complete, the resulting neutrino (background) datasets contained $716$ ($422$) collection plane selected hits in $552$ ($1970$) events.


\begin{table}[h]
\centering
\begin{tabular}{|c|c|c|c|}
\hline 
\multirow{2}{*}{Cut} & \multicolumn{3}{|c|}{Percent of Hits Remaining} \\ 
\cline{2-4}
 & Neutrino & Background & MC \\
\hline
Minimum Peak Height& $65$\% & $38$\% & $94$\%\\ 
\hline 
Maximum Peak Height & $58$\% & $37$\% & $84$\% \\ 
\hline 
Handscanning & $54$\% & $29$\% & $78$\% \\
\hline
Plane Matching & $24$\% & $10$\% & $54$\% \\
\hline 
\end{tabular} 
\caption{Impact of different cuts for collection plane hits. Cuts are applied sequentially. MC was simulated with no noise.}
\label{table:cuts}
\end{table}

Following this selection, we grouped signal hits into clusters and attempted a reconstruction of 
clusters' positions and energies. 
A cluster is defined as a collection of one or more signals on adjacent wires that occur within $40$ samples on these wires. This value was determined by examining a simulation of electrons with energies in the range of interest. If a cluster spans an unresponsive wire, each section was considered as a separate cluster. A total number of $553$, $319$ and $4537$ plane-matched clusters were reconstructed, yielding an average of $1.00$, $0.16$ and $1.12$ clusters per event in the selected neutrino, background and MC events, respectively. In neutrino events, most of the clusters ($75$\%) are composed of just one hit, $23$\% are two hit clusters, and only $2$\% are clusters with more than two hits.

\subsection{Position Reconstruction}
We reconstructed the 3D position of a cluster by matching the furthest upstream collection plane hit in a cluster to the furthest upstream induction plane hit in the matched cluster. This yielded a coordinate on the $yz$-plane. We then included the $x$-coordinate of the collection plane hit to obtain a 3D position and calculated the distance of each cluster with respect to the neutrino interaction vertex. While a cluster may span more than one wire in a plane, the distance traveled by the presumed Compton-scattered electron creating the cluster is negligible when compared to the distance from the vertex. 

\subsection{Charge to Energy Conversion}
To reconstruct the energy associated with each reconstructed cluster, first the measured pulse area (ADC $\times$ time) of each hit was converted to charge (number of ionization electrons) by an electronic calibration factor, then a lifetime correction was applied to account for ionization electron loss due to attachment on impurities in the liquid argon during drift, as described in ~\cite{CC1pion}. 

Calorimetric reconstruction in a LArTPC requires converting the collected charge to the original energy deposited in the ionization process. This requires applying a recombination correction which depends on charge deposition per unit length $dQ/dx$~\cite{main}. The low-energy photon-induced electrons in the present analysis result in just isolated hits or clusters of very few hits, not extended tracks, so the effective length of the electron track seen by a wire cannot be determined. 

\begin{figure*}[!t]
\includegraphics[scale=0.45,width=0.49\textwidth, trim = 0.2cm 0.3cm 0.2cm 1.0cm, clip]{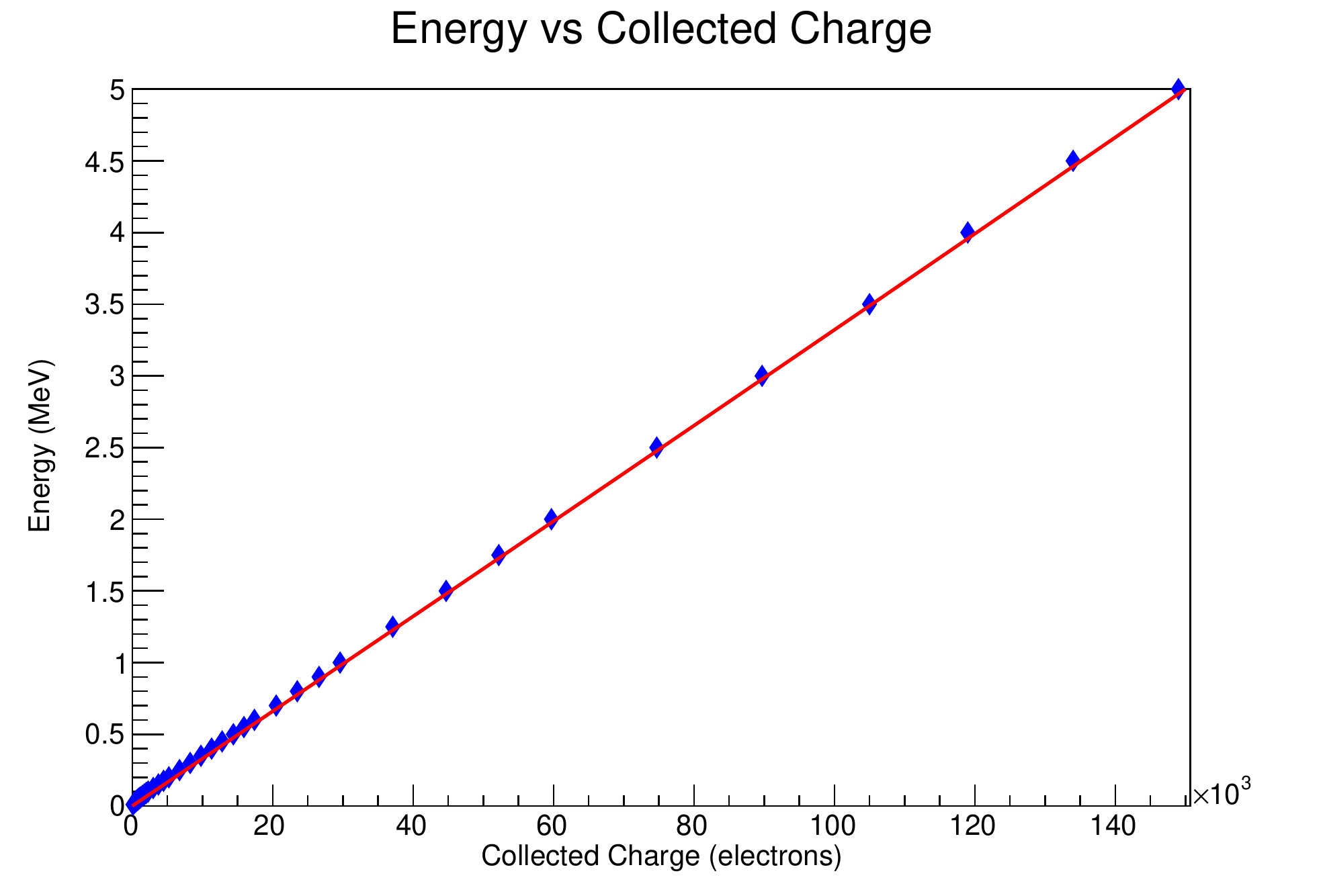}
\includegraphics[scale=0.4,width=0.49\textwidth, trim = 0.4cm 0.2cm 0.2cm 1.0cm, clip]{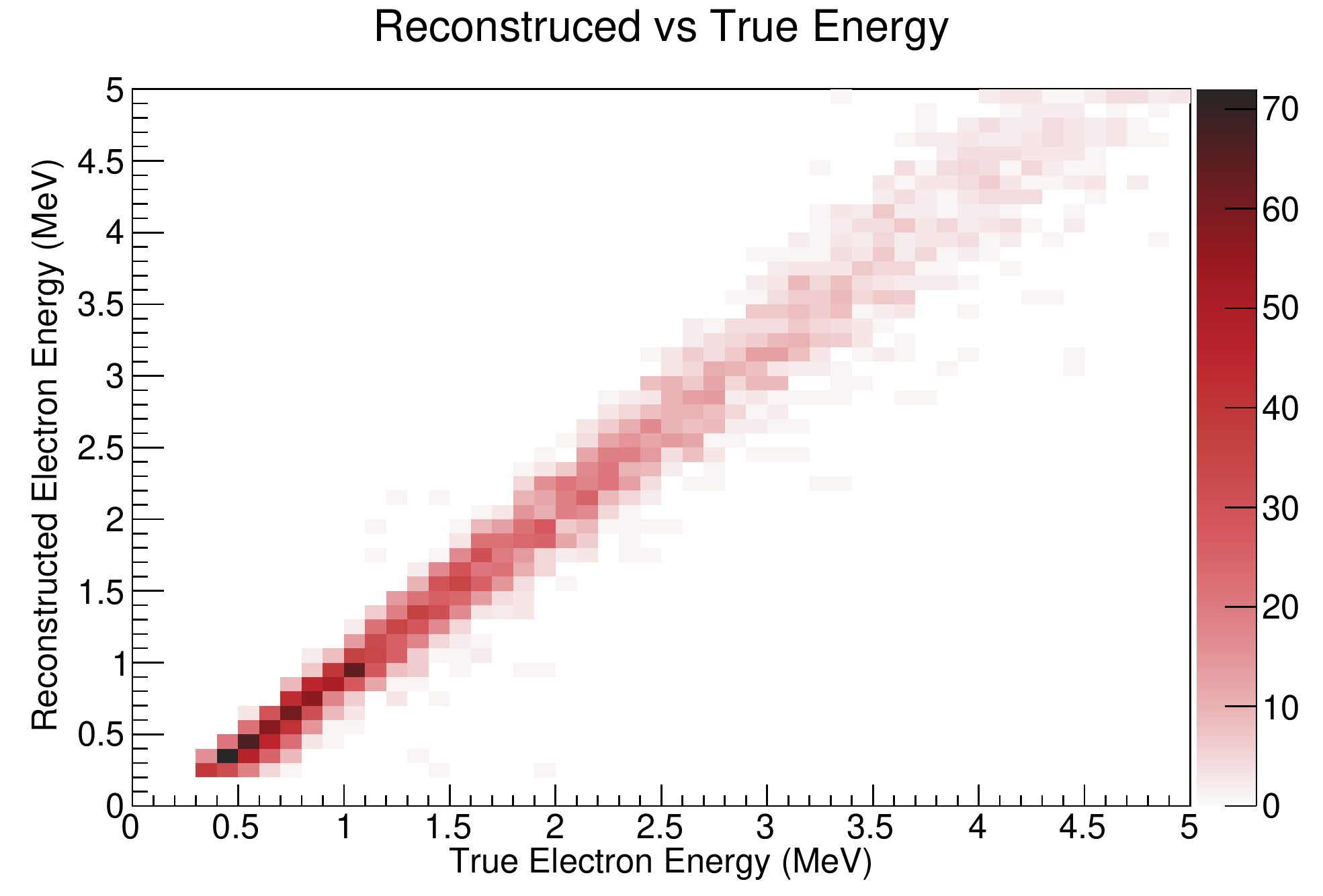}
\caption{Left: Energy deposited vs collected charge. Red curve indicates fit used to perform energy calculations from collected charge. 
Right: Reconstructed energy vs true electron energy using the charge method for a sample of simulated electrons with energies between $0$ and $5$ MeV. Events where the electron was not detectable are excluded.}
\label{fig:recotruecharge}
\end{figure*}

A different method to estimate the energy from the deposited charge which relies on the assumption that all hits passing cuts are due to electrons has been developed. The method uses the NIST table that provides the actual track length for electrons in LAr at given energies (ESTAR)~\cite{nist}, from $10$ keV to $1$ GeV. Using this table, we can thus approximate the deposited energy density $dE/dx$ by dividing the energy by the track length for each row in the table.  Using the Modified Box Equation~\cite{recombination} to model the recombination effect, we can calculate the expected $dQ/dx$ and by multiplying by the track length (i.e. $dx$), we obtain the expected amount of charge freed from ionization processes by an electron at a given energy, as shown in Fig.  \ref{fig:recotruecharge} (left). By using the result of a fit, also shown in the Figure, we can now convert collected charge from the individual hit to deposited energy. The total energy in a cluster is the sum of the deposited energy reconstructed for each individual hit forming the cluster.
To test the efficacy of this method, we applied it to a sample of GEANT4 simulated electrons propagating in LAr in the energy range of interest. Figure \ref{fig:recotruecharge} (right) indicates that it works well. We find a detection efficiency of $50$\% and energy resolution of $24$\% at $0.5$ MeV, and an efficiency of almost $100$\% and energy resolution of $14$\% at $0.8$ MeV.

\subsection{Systematic Uncertainties}
There are three primary sources of systematic uncertainty associated with hit and energy reconstruction in this analysis. As the electron lifetime varies 
between runs, we expect a variation and uncertainty in the number of near-threshold hits that are selected as signal. Despite having precise measurements of electron lifetime for all runs, we conservatively account for electron lifetime uncertainties by re-running FLUKA with a $\pm 25\%$ change in electron lifetimes; the resultant spread in reconstructed multiplicities and energies is treated as the systematic uncertainty from this source. A second systematic uncertainty arises from the choice of a true underlying functional form for the recombination correction. 
To account for this uncertainty, we consider reconstruction of simulated events using the unmodified Box Model as described in~\cite{recombination};
deviation from the default selection is treated as an uncertainty contribution from this source. Finally, there is a $3\%$ error associated with the utilized calorimetric calibration constants, which are fully correlated between all runs. Any multiplicity or energy variation arising from a $\pm 3\%$ shift in thresholds and reconstructed energies is treated as an uncertainty from this source. 
Systematic uncertainties in reconstructed positions are expected to be small and were not considered in this analysis.

\section{Results}
\label{sec:data}
\subsection{Comparison of Neutrino and Background Datasets}
Table \ref{table:signalvsbackground} shows a comparison of neutrino and background datasets. Comparing the different metrics leads to the conclusion that we have observed 
a statistically significant sample of neutrino-induced MeV-scale photons. 
Hit and cluster multiplicities are found to be significantly higher in the neutrino dataset than in the background dataset, with $1.30 \pm 0.07$ and $0.21 \pm 0.02$ hits per event, respectively. This difference corresponds to a $15 \sigma$ statistical excess of signal in the neutrino dataset. The higher neutrino dataset multiplicity is also accompanied by a larger per-event signal occupancy ($54 \pm 4\%$ in neutrino events versus $12 \pm 2\%$ in background events) and total signal energy per event ($1.1$ MeV in neutrino events versus $0.19$ MeV in background events). This can be interpreted as evidence of neutrino-induced MeV-scale energy depositions.  

\subsection{Comparison to MC Simulations}
\begin{figure}[t]
\includegraphics[scale=0.45, trim = 0.5cm 0.3cm 0.3cm 1.1cm, clip]{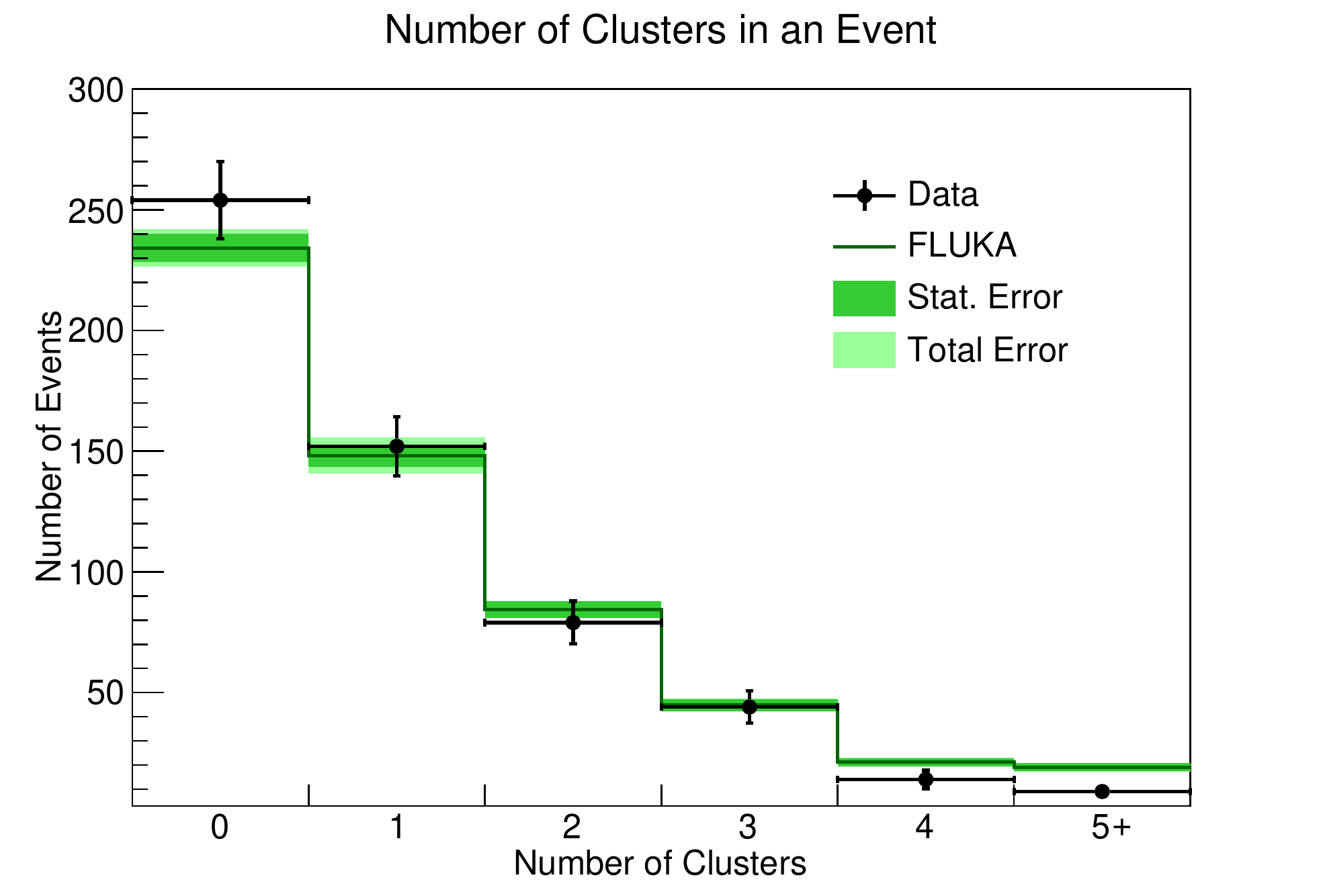}
\caption{Cluster multiplicity for neutrino data and FLUKA MC events. Data points include statistical error. Dark green line indicates FLUKA prediction with data-driven background added (see text). Dark green shaded area is statistical error in FLUKA, overlaid on total error (statistical + systematic) for FLUKA in light green shading. MC is normalized to the number of neutrino data events.}

\label{fig:nClus}
\end{figure}

\begin{figure}[t]
\includegraphics[scale=0.45, trim = 0.5cm 0.2cm 0.3cm 1.1cm, clip]{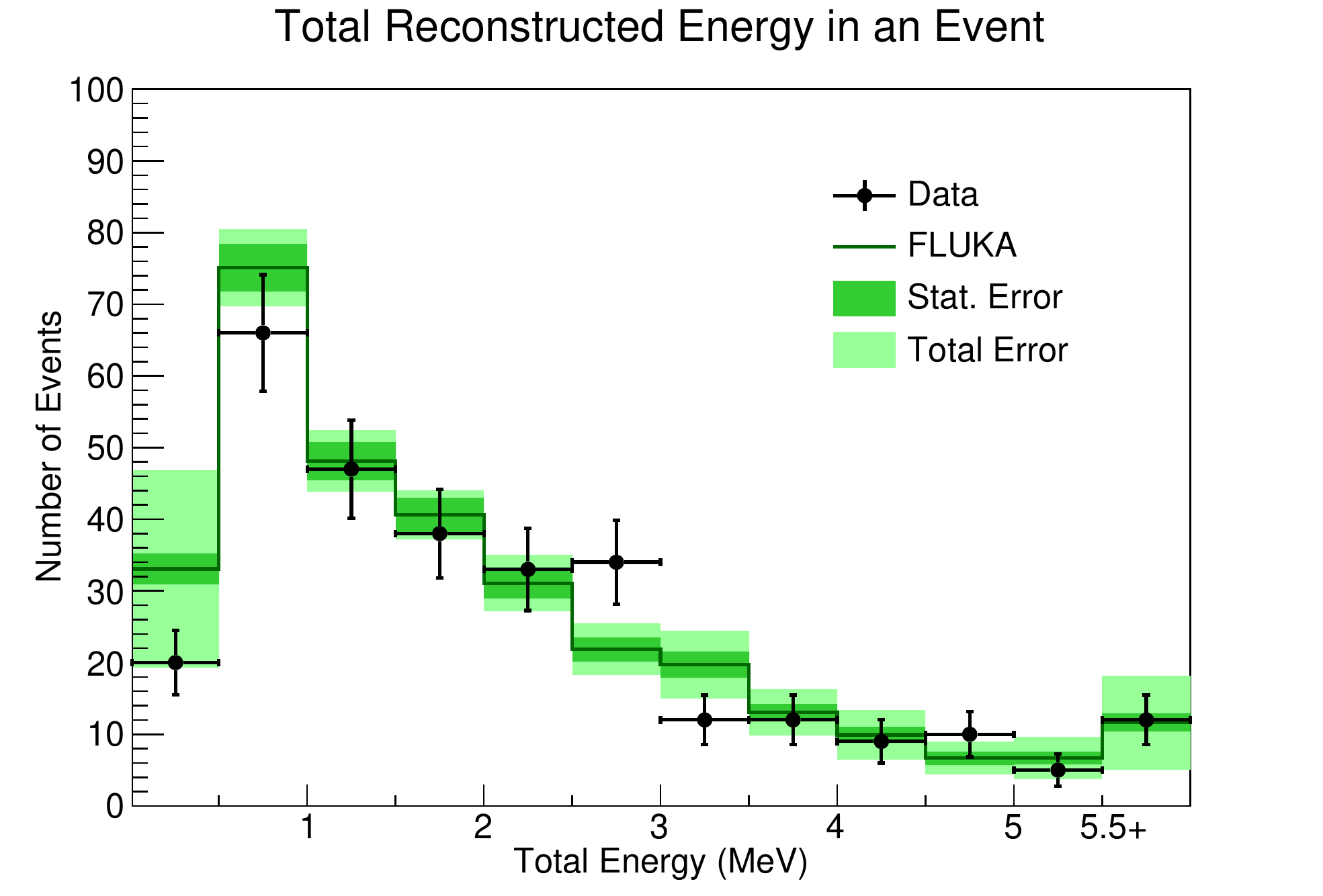}
\caption{Total signal reconstructed energy in an event for neutrino data and FLUKA MC events. Events with no reconstructed energy are not included. Data points include statistical error. Dark green line indicates FLUKA prediction with data-driven background added (see text). Dark green shaded area is statistical error in FLUKA, overlaid on total error (statistical + systematic) for FLUKA in light green shading. MC is normalized to the number of neutrino data events.}
\label{fig:totalE}
\end{figure}

A comparison of reconstructed per-event signal multiplicity and total signal energy for data and FLUKA MC simulation are shown in Figs. \ref{fig:nClus} and \ref{fig:totalE}, respectively.

\begin{table}
\begin{tabular}{|l|l|l|}
\hline 
Metric & Neutrino Data & Background\\ 
\hline 
Number of hits per event & $1.30$ & $0.21$ \\ 
\hline 
Number of clusters per event & $1.00$ & $0.16$ \\ 
\hline 
Average total signal energy & \multirow{2}{*}{$1.11$} & \multirow{2}{*}{$0.19$} \\ 
in an event (MeV) & & \\
\hline 
Percent of events with & \multirow{2}{*}{$54\%$} & \multirow{2}{*}{$12\%$} \\
at least one signal hit & & \\
\hline
Average cluster distance & \multirow{2}{*}{$22.4$} & \multirow{2}{*}{$-$} \\ 
from vertex (cm) & & \\
\hline 
\end{tabular} 
\caption{Comparison of neutrino and background datasets when examining hits passing all cuts. The difference in the first four metrics indicates neutrino-induced MeV-scale activity is visible.}
\label{table:signalvsbackground}
\end{table}

In both data and MC, around half of the events have no signal clusters, as expected based on the small ArgoNeuT detector size and the previously-mentioned sizable number of predicted product nuclei in the ground-state. Overall, there is good agreement between data and FLUKA MC predictions. We find a $\chi^2 / \text{ndf}$ of $7.81/12$ (p-value  $0.80$) for the total reconstructed energy distributions, and a $\chi^2 / \text{ndf} = 12.6/6$ (p-value $0.05$) for the cluster multiplicity distribution. Thus, we observe that FLUKA, which incorporates low-level nuclear processes that result in the production 
of MeV-scale energy depositions following interactions of GeV-scale neutrinos in liquid argon, agrees well with the data. We observe that the largest contributor to the $\chi^2$ between the data and MC multiplicity distributions is the difference in high-multiplicity events. The modest excess in MC, which spreads over multiple reconstructed energy bins, could be indicative of flaws in the hit selection process, or of imperfections in models or  libraries utilized by FLUKA. This feature can be better examined in future high-statistics studies in larger LArTPCs. Finally, we notice a dip in the first bin in Fig. \ref{fig:totalE}, due to detector thresholding, which can vary in data from event to event due to different electron lifetime values.

\begin{table}
\begin{tabular}{|l|l|l|l|}
\hline 
Metric & De-excitation & Neutron & Total\\ 
\hline 
Number of hits per event & $0.48$ & $0.98$ & $1.46$ \\ 
\hline 
Number of clusters per event & $0.35$ & $0.77$ & $1.12$ \\ 
\hline 
Average event energy (MeV) & $0.41$ & $0.76$ & $1.17$ \\ 
\hline 
Average cluster energy (MeV) & $1.18$ & $0.98$ & $1.04$ \\ 
\hline 
Average hit energy (MeV) & $0.86$ & $0.77$ & $0.80$ \\ 
\hline 
Average cluster distance & \multirow{2}{*}{$15.7$} & \multirow{2}{*}{$23.4$} & \multirow{2}{*}{$21.0$} \\ 
from vertex (cm) & & & \\
\hline 
\end{tabular} 
\caption{Relative contributions of de-excitation and neutron-produced photon components in FLUKA MC.}
\label{table:mcstats}
\end{table}

\begin{figure}[h]
\includegraphics[scale=0.45, trim = 0.5cm 0.3cm 0.3cm 1.1cm, clip]{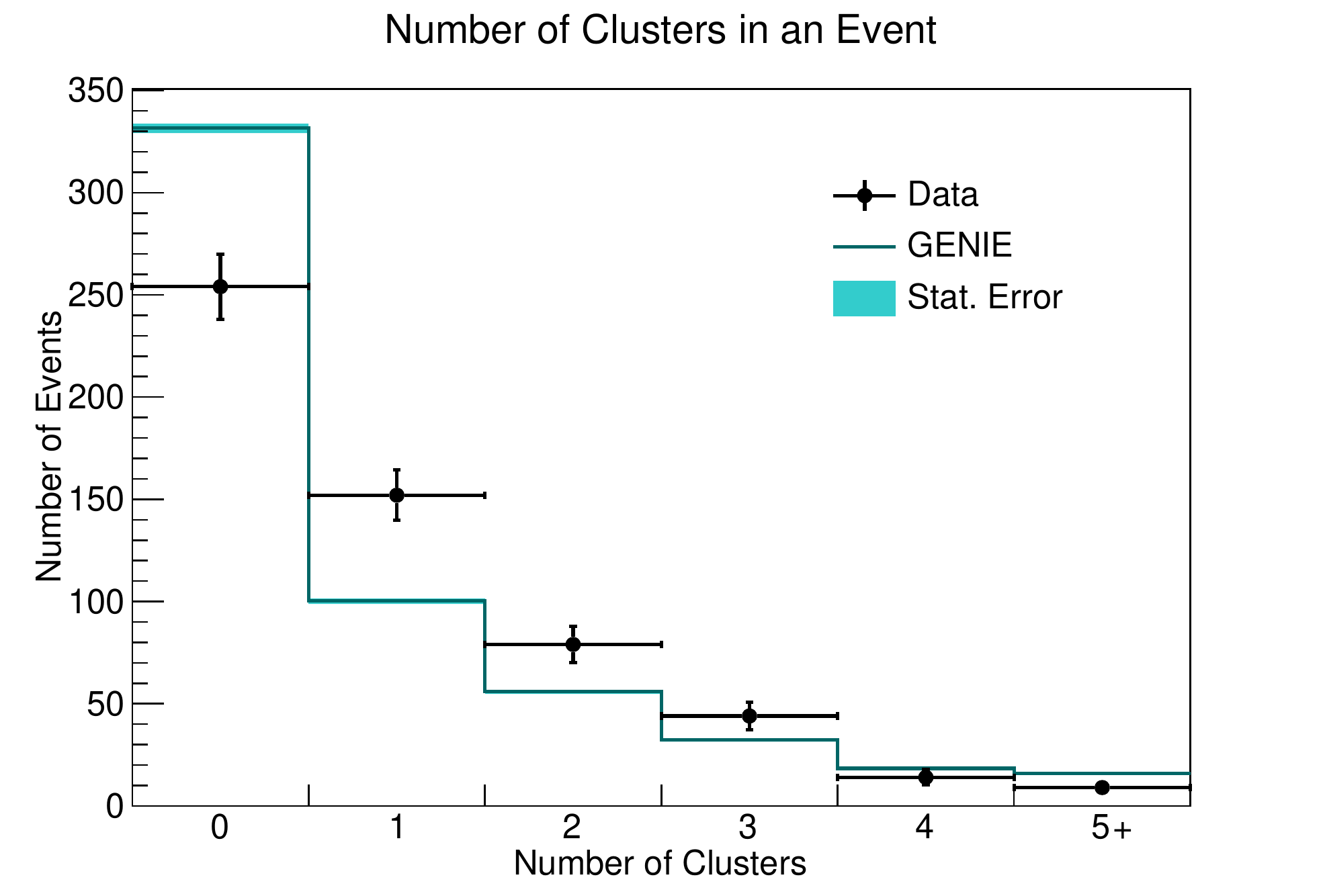}
\caption{Distribution of cluster multiplicity for neutrino data and GENIE events. Data points include statistical error. Dark blue indicates GENIE prediction (no de-excitation photons). Light blue shaded area indicates statistical error for GENIE prediction. MC is normalized to the number of neutrino data events.}
\label{fig:nClusgenie}
\end{figure}

Both components, de-excitation photons and photons produced by interactions of final-state neutrons on argon, are needed to have data-MC agreement. If de-excitation photons are removed from FLUKA distributions, we obtain a $\chi^2 / \text{ndf} = 82.6/12$ for reconstructed energy and  $\chi^2 / \text{ndf} = 93.8/6$ for the cluster multiplicity. If neutron-produced photons are removed, we obtain  $\chi^2 / \text{ndf} = 194/12$ and $\chi^2 / \text{ndf} = 197/6$ for these same distributions, respectively. To confirm this, we also compared ArgoNeuT data with a GENIE MC simulation~\cite{genie}; existing user interfaces allowed for easy generation of GENIE final states within the LArSoft framework. The same event selection and reconstruction procedure as in FLUKA was applied to GENIE events. 
As an example, a comparison of reconstructed multiplicity is shown in Fig. \ref{fig:nClusgenie}. The $\chi^2 / \text{ndf} $ is $57.9/6$. This disagreement is attributed to the lack of de-excitation photons in the GENIE simulation of neutrino-argon interactions.


\begin{figure}[h]
\includegraphics[scale=0.45, trim = 0.5cm 0.3cm 0.3cm 1.2cm, clip]{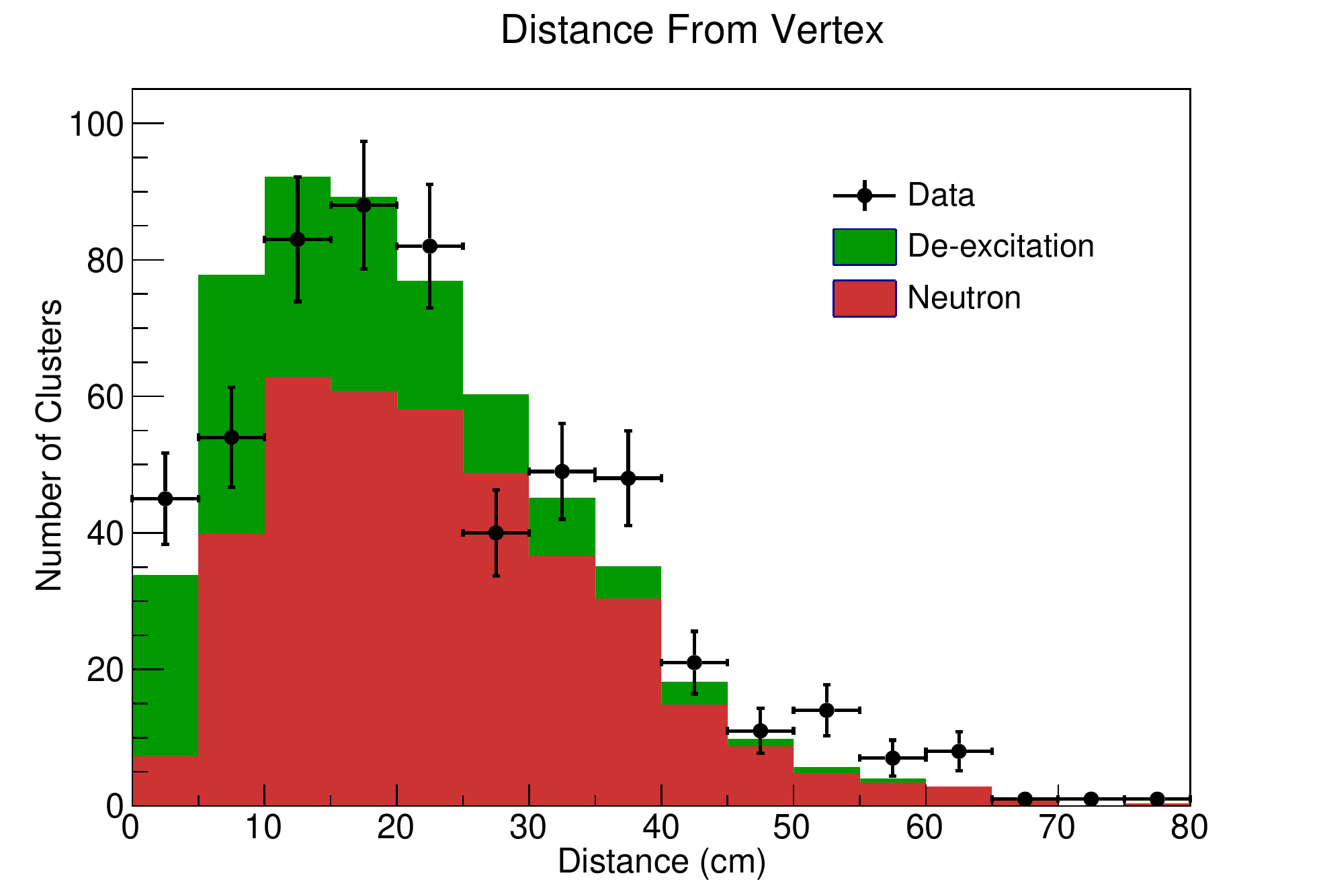}
\caption{Distributions of cluster position with respect to the neutrino interaction vertex in neutrino data and FLUKA MC events. Data includes statistical error. Green indicates the contribution of photons from de-excitation of the final-state nucleus. Red indicates the contribution of photons from inelastic neutron scattering. MC is area normalized to data.}
\label{fig:distance}
\end{figure}

These results indicate that the observed MeV-scale signals in ArgoNeuT contain both de-excitation and neutron-produced photons. The contribution of each of these sources to the total activity in an event as given by the FLUKA simulation is shown in Table \ref{table:mcstats}. We find that we cannot distinguish between the two sources of photons by examining the energy of a hit or cluster alone, but we do see a difference in the distance of a cluster with respect to the neutrino interaction vertex. The distribution of these distances is seen in Fig. \ref{fig:distance}.
Photons produced by de-excitation of the final-state nucleus tend to be concentrated at lower distances, while photons produced by inelastic neutron scattering dominate at higher distances. 

\section{Conclusion}
\label{sec:conclusion}
The ability to reconstruct activity at the MeV scale in a LArTPC is crucial for future studies of supernova, solar, and beam neutrino interactions. In addition, studies of low scale new physics scenarios, such as millicharged particles, light mediators, and inelastic scatterings with small splittings (see e.g. Refs.~\cite{Gninenko:2012rw, Magill:2018tbb, Bertuzzo:2018ftf}), could invaluably profit from such low energy reconstruction.  By studying low-energy depositions produced by photons in ArgoNeuT neutrino interactions and comparing to simulation, we have shown that such a reconstruction is possible. Performing this study required the creation of new techniques for low-energy LArTPC reconstruction. By reconstructing photons produced by nuclear de-excitation and inelastic neutron scattering, we have extended the LArTPC's range of physics sensitivity down to the sub-MeV level, reaching a threshold of $0.3$ MeV in this analysis. This range now spans more than three orders of magnitude, up to the GeV level. 

In our study of low-energy depositions in $552$ ArgoNeuT neutrino events, we found $553$ clusters with an average of $1.30 \pm 0.07$ hits per event and an average energy of $1.11 \pm 0.16$ MeV per event. Signal cluster multiplicities in neutrino events outnumbered those in nearby background events, establishing a clear neutrino-based origin for these MeV-scale features. These and other cluster properties matched those predicted for photons due to inelastic neutron scattering and de-excitation of the final-state nucleus in FLUKA using its model of nuclear physics processes at the MeV-scale. Removal of either of these event classes significantly worsens the level of data-simulation agreement.


This analysis represents the first-ever reported detection of de-excitation photons or final-state neutrons produced by beam neutrino interactions in argon. Both of these particle classes could provide valuable new avenues of investigation for physics reconstruction in LArTPCs.  Reconstruction of MeV-scale neutron-produced features may enable some level of direct reconstruction of final-state neutron energies or multiplicities, which would provide a valuable new handle on one of the dominant expected differences between neutrino and antineutrino interactions in liquid argon.  Precise reconstruction of de-excitation photon multiplicities and energies will improve overall reconstruction of neutrino energies, particularly for those at lower energies, such as supernova and solar neutrinos.  
Future MC studies and higher-statistics datasets from future large LArTPCs will provide additional understanding of the value of these MeV-scale features. 

\section {Acknowledgements}
This manuscript has been authored by Fermi Research Alliance, LLC under Contract No. DE-AC02-07CH11359 with the U.S. Department of Energy, Office of Science, Office of High Energy Physics. We gratefully acknowledge the cooperation of the MINOS Collaboration in providing their data for use in this analysis. We wish to acknowledge the support of Fermilab, the Department of Energy, and the National Science Foundation in ArgoNeuT's construction, operation, and data analysis. We also wish to acknowledge the support of the Neutrino Physics Center (NPC) Scholar program at Fermilab, ARCS Foundation, Inc and The Royal Society (United Kingdom).

This material is based upon work supported by the U.S. Department of Energy, Office of Science, Office of Workforce Development for Teachers and Scientists, Office of Science Graduate Student Research (SCGSR) program. The SCGSR program is administered by the Oak Ridge Institute for Science and Education (ORISE) for the DOE. ORISE is managed by ORAU under contract number DE-SC0014664.
\bibliographystyle{apsrev4-1}
\bibliography{references}

\end{document}